\documentclass{appolb}
\usepackage{graphicx}
\usepackage{lineno}

\begin{document}
\title{Event-by-event multiplicity fluctuations in Pb-Pb collisions in ALICE%
\thanks{Presented at WPCF 2015}
}
\author{maitreyee mukherjee for the alice collaboration
\address{Variable Energy Cyclotron Centre, Kolkata, India\\
E-mail : maitreyee.mukherjee@cern.ch\\
}
}
\maketitle

\begin{abstract}
Fluctuations of various observables in heavy-ion collisions at ultra-relativistic energies have been extensively studied as they provide important signals regarding the formation of a Quark-Gluon Plasma (QGP). Because of the large number of produced particles in each event, a detailed study of event-by-event multiplicity fluctuations has been proposed as one of the signatures of the phase transition. In addition, the understanding of multiplicity fluctuations is essential for other event-by-event measurements.
In the present work, we have calculated the scaled variance ($\omega_{\rm ch}=\sigma^{\rm 2} / \mu$) of the charged-particle multiplicity distributions as a function of centrality in Pb-Pb collisions at LHC energies. Here, $\mu$ and $\sigma$ denote the mean and the width of the multiplicity distributions, respectively. The trend of scaled variances as a function of centrality is presented and discussed. Volume fluctuations play an important role while measuring the multiplicity fluctuations, which are also discussed. The results are expected to provide vital input to theoretical model calculations.
\end{abstract}
\PACS{25.75.-q,25.75.Gz,25.75.Nq,12.38.Mh}


\section{Introduction}
The ALICE detector~\cite{alice} at the Large Hadron Collider (LHC) is within the crossover region in the QCD-phase diagram, i.e, in the region of very low net-baryon-density and high temperature. The main aim of the ALICE experiment is to study the strongly interacting matter at this extreme energy density and temperature, where the Quark-Gluon Plasma (QGP) is expected. The understanding of deconfinement and the chiral symmetry restoration are other vital studies.\\Event-by-event fluctuations, such as, fluctuations of mean transverse momentum, net charge, multiplicity, higher moments, particle ratios, etc., basically provide information on the details of the dynamics of the colliding system. Large numbers of produced particles in each event make these studies possible.
\section{Motivation of Multiplicity Fluctuation Analysis}
The QCD phase transition can manifest itself by characteristic behaviour of the observables which vary dramatically from one event to another. Following the properties of the Grand Canonical Ensemble (GCE)~\cite{jeon}, the variance ($\sigma^{\rm 2}$) of the charged-particle multiplicity distribution is connected to a thermodynamic quantity, which is the isothermal compressibility ($k_{\rm T}$) of the produced system as~\cite{Phenix},
\begin{equation}
\sigma^{\rm 2} = \frac{k_{\rm B}T\mu^{\rm 2}}{V}k_{\rm T}
\end{equation}
where $\mu$ is the mean multiplicity and $k_{\rm B}$ is Boltzmann's constant. Therefore, the multiplicity fluctuation expressed in terms of the scaled variance ($\omega_{\rm ch}=\sigma^{\rm 2}/\mu$) is directly proportional to $k_{\rm T}$. $k_{\rm T}$ increases by ~an order of magnitude close to the QCD critical point (CP), following a power-law scaling with critical exponent $\gamma$, which basically has the identical value for the same universality class of systems. 
\begin{equation}
k_{\rm T} \propto \left( \frac{T-T_{\rm c}}{T_{\rm c}} \right)^{-\rm \gamma}
\end{equation}
Thus, it is also possible to group the systems into universality classes (this is a grouping of systems in the nature)~\cite{Phenix}.
\subsection {Motivation for Multiplicity Fluctuation Analysis in ALICE}
Dynamical fluctuations (other than the statistical fluctuations and fluctuations in the number of participants) and multiplicity distribution measurements provide constraints on the particle production models. Measurements at vanishing $\mu_{\rm B}$ set the scale of the theoretical calculations. Besides, at LHC energies, values of Bjorken-x below $10^{-4}$ can be accessed, where initial state effects can be studied with multiplicity measurements~\cite{cgcreview}. Model calculations with CGC initial energy distributions have shown that experimental multiplicity distributions from $d+Au$ collisions at RHIC are better explained if multiplicity fluctuations are included~\cite{dumitru}.

\section{Analysis Details}
The analysis has been performed using 14 M minimum bias Pb-Pb events. The data for the analysis were taken at $\sqrt{s_{\rm NN}}=2.76$~TeV. The following detectors have been used for this analysis : Silicon Pixel Detectors (SPD) for vertex determination, V0 (forward detectors) for centrality selection and the Time Projection Chamber (TPC) for charge-particle selection. Charged particles are selected within $0.2 < p_{\rm T} < 2.0~{\rm GeV}/c$ and $|\eta| < 0.8$.
\subsection{Centrality Selection}
Centrality is selected using the charged-particle minimum-bias distribution of the V0 amplitude. The centrality percentile is determined by fitting the minimum-bias distribution with the Glauber Model, thus connecting the measured multiplicity to the number of participating nucleons ($N_{\rm part}$) and number of binary nucleon-nucleon collisions ($N_{\rm coll}$). Due to the non-uniformity in charged-particle distributions, the Centrality Binwidth Effect (CBW) arises. The prescription is to divide one centality bin into smaller bins and weight the moments as,
\begin{equation}
X = \frac{\sum_{i}  n_{\rm i}X_{\rm i}}{\sum_{i}n_{\rm i}},
\end{equation}
where the index $i$ runs over each multiplicity bin, 
$X_{\rm i}$ represents various moments for the $i$-th bin, and 
$n_{\rm i}$ is the number of events in the $i$-th multiplicity bin. 
$\sum_{i}n_{\rm i} = N$ is the total number of events in the centrality bin ~\cite{nihar}.
Final results will be shown for $5\%$~centrality bins, after applying the bin width correction.
\subsection{Estimation of volume fluctuations}
The scaled variance depends on the fluctuations of $N_{\rm part}$. According to Heiselberg~\cite{heiselberg}, considering the participant model, the scaled variance can be written as,
\begin{equation}
\omega_{\rm ch}=\omega_{\rm n}+\langle n \rangle \omega_{N_{\rm part}}
\end{equation}
where $\langle n \rangle$~is the mean multiplicity of hadrons from a nucleon-nucleon source and $\omega_{N_{\rm part}}$~is the scale of the fluctuation in $N_{\rm part}$. 
\begin{figure}[htb]
\begin{center}
\includegraphics[scale=0.4]{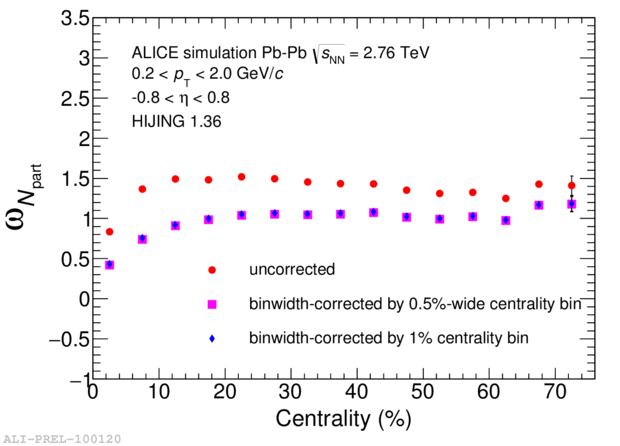}
\caption{Volume fluctuations in Pb-Pb collisions at $\sqrt{s_{\rm NN}}=2.76$~TeV for $0.2 < p_{\rm T} < 2.0~{\rm GeV}/c$~and~$-0.8\le \eta \le0.8$, as obtained by a HIJING simulation.
}
\label{fig1}
\end{center}
\end{figure}
Figure~\ref{fig1} shows results for $\omega_{N_{\rm part}}$~from HIJING. It is observed that by choosing narrow bins in centrality, fluctuations in $N_{\rm part}$ are minimized. In this case, the scale of the fluctuation in $N_{\rm part}$ is close to unity. 

\subsection{Correction for detector inefficiency}
Detector efficiency ($\epsilon$) generally is defined as the fraction of the number of accepted tracks from primary particles to the number of all primary particles. For ALICE, $\epsilon$~is not a constant, rather it has a non-flat $p_{\rm T}$ dependence. To correct for these local efficiency effects (following the prescription given in~\cite{bzdak}), 9 $p_{\rm T}$ intervals have been used for each centrality, and for each such bin, $\epsilon$-values are calculated and the numbers of charged particles are counted. Thus, the corrected factorial moments are calculated, from which we get efficiency-corrected values for $\mu$, $\sigma$~and~$\omega_{\rm ch}$.
\begin{figure}[hbtp]
\begin{center}
\includegraphics[scale=0.4]{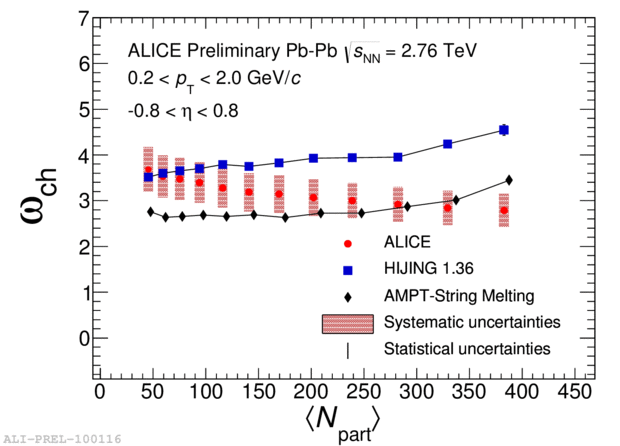}
\caption{Comparison between data and model results. HIJING and AMPT (with string melting option and with 200 k generated events)  models are used here. Statistical errors are within the data points.}
\label{fig2}
\end{center}
\end{figure}

\section{Results}
In Figure~\ref{fig2}, the result for the scaled variance ($\omega_{\rm ch}$) is shown as a function of the number of participating nucleons ($N_{\rm part}$). The systematic uncertainty for $\omega_{\rm ch}$~is~$\sim 13\%$. The main sources of the systematic uncertainties are the resolution effect, changing track and vertex selection criteria, different magnetic field polarity, tracking efficiency and data cleanup. We observe that the scaled variance is decreasing slowly from
peripheral ($\omega_{\rm ch} \sim$~3.6) to central ($\omega_{\rm ch} \sim$~2.8) collisions. Neither HIJING, nor AMPT (with string melting option) can describe the trend from data, though the results from the models are of comparable values as from data.\\Further studies are ongoing on the effect of acceptance on the observables.


\end{document}